\documentclass[preprint,preprintnumbers,amsmath,amssymb,nofootinbib,superscriptaddress]{revtex4}

\usepackage{graphicx}% Include figure files
\usepackage{dcolumn}% Align table columns on decimal point
\usepackage{bm}% bold math

\begin{document}

%\preprint{}

\vspace*{1.5cm}

\title{Quantum Criticality in Einstein-Maxwell-Dilaton Gravity}

\author{Wen-Yu Wen}
\email{steve.wen@gmail.com}
\affiliation{California Institute of Technology \\
Pasadena, CA 91125, USA}
\affiliation{Department of Physics and Center for Theoretical Sciences \& \\ 
Leung Center for Cosmology and Particle Astrophysics,\\
National Taiwan University, Taipei 106, Taiwan}

%\date{\today}

\begin{abstract}
We investigate the quantum Lifshitz criticality in a general background of Einstein-Maxwell-Dilaton gravity.  In particular, we demonstrate the existence of critical point with dynamic critical exponent $z$ by tuning a nonminimal coupling to its critical value.  We also study the effect of nonminimal coupling and exponent $z$ to the Efimov states and holographic RG flow in the overcritical region.  We have found that the nonminimal coupling increases the instability for a probe scalar to condensate and its back reaction is discussed.  At last, we give a quantum mechanics treatment to a solvable system with $z=2$, and comment for generic $z>2$.
\end{abstract}

%\pacs{}
%\keywords{}

\maketitle

%%%%%%%%%%%%%%%%%%%%%%%%%%%%%%%%%%%%%%%%%%%%%%%%%%%%%%%
\section{Introduction}
%%%%%%%%%%%%%%%%%%%%%%%%%%%%%%%%%%%%%%%%%%%%%%%%%%%%%%%

Critical points, across which a continuous phase transition happens, are interesting and important for their universality, meaning that they can be simply classified according to very few critical exponents.  In particular, a quantum Lifshitz point, where fluctuation is driven by zero point energy and characterised by anisotropic scaling of space and time, might be realized in some antiferromagnetic matters with strongly correlated electrons.  As an alternative to the conventional lattice approach toward nonperturbative computation, application of AdS/CFT correspondence, originally proposed as a duality between strings in weakly curved AdS space and operators in strongly coupled super Yang-Mills\cite{Maldacena:1997re,Gubser:1998bc,Witten:1998qj}, to quantum critical points in strongly coupled systems has demonstrated some interesting results\cite{Faulkner:2009wj}.   In this paper, we would like to study the quantum criticality in a more general background of Einstein-Maxwell-Dilaton gravity.  From the theoretical perspective, descending from the (super)gravity in higher dimensional spacetimes, it is very common to find a gravity system in lower dimensions couple nonminimally to a number of dilatons, gauge fields, higher ranked tensor and form fields.  From the practical viewpoint, there are at least two advantages along this line of generalization, which will become clear later: 

\begin{enumerate}
	\item The nonzero dilaton field supports the Lifshitz-like scaling as the isometry of background metric, such that quantum Lifshitz point becomes accessible.
	\item	The nonminimal couple between a probe scalar and the Maxwell field, as well as the direct couple between two scalars, provide tunable parameters in addition to scalar masses.  Since the mass of a scalar will map to the conformal dimension of corresponding condensate\footnote{For this statement to be true, we have implied that our background is asymptotically $AdS_4$ at infinity.}, we are able to approach the quantum critical point while keeping the scaling dimension unaltered.
\end{enumerate}

This paper is organized as follows.  The gravity model and its probe limit is introduced in the section II.  The effect of nonminimal couple to the quantum criticality at $AdS_2$, as a special case, will be discussed in the section III.  The quantum criticality at Lifshitz point for generic critical exponent is discussed in the section IV and BKT phase transition in the section V.  We will discuss the solution beyond the probe limit in the section VI and discussion and  comments in the last section.  In the appendix we give a quantum mechanics treatment for solvable case $z=2$.

\section{The gravity model and its probe limit}
We will consider the following Lagrangian as a generalized model of Einstein-Maxwell-Dilaton gravity\cite{Aprile:2010yb}:
\begin{equation}
2\kappa_G^2 (-g)^{-1/2}{\cal L} = R + \frac{6}{L^2} - G(\psi,\chi) F^{\mu\nu}F_{\mu\nu} -|D_{\mu}\psi|^2 - |D_{\mu}\chi|^2 - V(\psi,\chi),
\end{equation}
where $\psi$ and $\chi$ are complex scalars carrying charges $q_\psi$ and $q_\chi$ under Maxwell field $F_{\mu\nu}=\partial_{[\mu} A_{\nu]}$.  Since the phases of scalars are irrelevant to our discussion on uniform condensate, it is consistent to set them to zero.  Similar constructions, which can be seen as special limits of this model, have been useful to simulate various condensed matter systems.  To mention a few: a single charged scalar can be used to describe the superconductor\cite{Hartnoll:2008vx}, a neutral scalar probed in the charged black hole can be used to model the antiferromagnetic state\cite{Iqbal:2010eh}, and competing of two (charged) scalars was first attempted in \cite{Basu:2010fa} for mixed magnetic and superconducting states.  In this paper, we adopt a generalized background where additional function $G$ is introduced to engineer possible interaction between scalars and Maxwell field, other than the minimal coupling via covariant derivative $D_\mu$.  Some new features of this generalization have been observed in \cite{Aprile:2010yb} where, for instance, the critical temperature becomes tunable and phase transition other than second order can be engineered.  We will choose a specific form of $G$ and $V$ for a toy model:
\begin{eqnarray}\label{model}
&&G(\chi) = 1 + \kappa |\chi|^2,\nonumber\\
&&V(\psi,\chi) = m_\psi^2 |\psi|^2 + m_\chi^2 |\chi|^2 + \eta |\psi|^2|\chi|^2. 
\end{eqnarray}

We would like to study the limit similar to that in \cite{Iqbal:2010eh}, where the boundary theory is at zero temperature and finite density.  To achieve this, we take a probe limit of scalar field $\chi$ such that it decouples from the rest of the fields\footnote{Notice that this limit is different from the usual probe limit for Einstein-Maxwell model where both scalar and vector fields are decoupled from the gravity sector upon sending $q\to \infty$ after scaling down both $\psi$ and $A$ by a factor of $q$.}.  The above mentioned action will break down into two pieces: the background in its IR region ($u\to \infty$), supported by the Maxwell field and constant scalar $\psi$, admits a geometry respecting the Lifshitz scaling of critical exponent $z$\cite{Gubser:2009cg}:
\begin{eqnarray}
&& ds^2  = -(\frac{L_0}{u})^{2z}dt^2 + \frac{L_0^2}{u^2}(d\vec{x}^2 + du^2),\nonumber\\
&& A_t = \sqrt{2-\frac{2}{z}} (\frac{L_0}{u})^z, \qquad \psi = \psi_0
\end{eqnarray}
where the constant $\psi_0$, charge $q$, and radius of curvature at IR $L_0$ are determined by a pair $(z,m_\psi^2)$ for $m_\psi^2>0$:
\begin{equation}\label{sol_z}
\psi_0 = \frac{\sqrt{2(z-1)}}{m_\psi L_0},\qquad q_\psi^2 = \frac{zm_\psi^2}{2(z-1)}, \qquad L_0 = L \sqrt{\frac{(z+1)(z+2)}{6}}.
\end{equation}
Given such an IR geometry with $z>1$, it is unclear whether a corresponding UV solution can be exactly constructed\footnote{We remark that the charged dilatonic black hole and brane have been numerically constructed, for example, in \cite{Goldstein:2009cv,Bertoldi:2009dt}, where the near horizon geometry exhibits the Lifshitz spacetime and it becomes $AdS_4$ at asymptotical infinity.}.  One well known example is given by the extremal RN black hole in $AdS_4$\footnote{Here we already rescale the horizon at $u=1$.  The extremal limit is obtained for $\alpha = 1$.}: 
\begin{equation}
ds_{UV}^2 = \frac{L^2}{u^2}(-f(r)dt^2+\frac{du^2}{f(r)}+d\vec{x}^2),\qquad f(r) = 1 - (1+3\alpha) u^3 + 3\alpha u^4.
\end{equation}
It flows to $AdS_2\times R^2$ in the IR region, which corresponds to a RG flow in the boundary field theory from a UV fixed point with exponent $z=1$ to an IR one with $z\to \infty$. The near horizon solution is given by
\begin{equation}\label{ads2_metric}
ds_{IR}^2 = \frac{L_0^2}{u^2}(-dt^2+du^2)+\frac{d\vec{x}^2}{L^2},
\end{equation}
and 
\begin{equation}\label{ads2_at}
A_t = \frac{L_0^2\mu}{u}, \qquad \psi = 0,
\end{equation}
where the curvature radius of $AdS_2$ can be related to that of $AdS_4$ via
\begin{equation}
L_0^2=\frac{L^2}{6}.
\end{equation}
On the other hand, the probe action for scalar field $\chi$ now reads:
\begin{equation}\label{probe_L}
2\kappa_G^2(-g)^{-1/2}{\cal L} = - \frac{L^2}{4} \kappa|\chi|^2F^{\mu\nu}F_{\mu\nu}-m_\chi^2|\chi|^2-\eta|\psi|^2|\chi|^2-|\partial_\mu \chi - iq_\chi A_{\mu}\chi|^2.
\end{equation}
The condition to have instability in its IR region (for $\chi$ to condensate) would depend on not only the pair $(z,m_\psi)$, but also $(\kappa,\eta)$, representing a nontrivial interaction among $\chi$, Maxwell field $F$ and background scalar $\psi$.

%%%%%%%%%%%%%%%%%%%%%%%%%%%%%%%%%%%%%%%%%
\section{Quantum Criticality at $AdS_2$}
%%%%%%%%%%%%%%%%%%%%%%%%%%%%%%%%%%%%%%%%%
  
As a warm up, we will revisit the quantum criticality at $AdS_2$ before going for generic $z$.  A detail treatment for a minimal coupled scalar was given in \cite{Faulkner:2009wj}, so here we only highlight the difference.  Let us make a Fourier transform of the scalar field along $(t,\vec{x}$ directions:
\begin{equation}
\chi(u,t,\vec{x}) = \int{\frac{d\omega d^2k}{(2\pi)^3}}\chi(u,\omega,\vec{k})e^{-i\omega t +\vec{k}\cdot\vec{x}}, \quad |\vec{k}|\equiv k.
\end{equation}
Now consider the metric of $AdS_2\times R^2$ with a constant electric field, obtained in (\ref{ads2_metric}) and (\ref{ads2_at}).  The equation of motion reads:
\begin{eqnarray}\label{eqn_inner}
&&-\partial_u^2 \chi + \left[ \frac{m_{eff}^2L_0^2}{u^2} - (\omega+\frac{\mu q_\chi L_0}{u})^2 \right]\chi = 0 ,\nonumber\\
&&m_{eff}^2 \equiv m_\chi^2 + k^2L^2 - 6\kappa \mu^2.
\end{eqnarray}

\begin{figure}[tbp]
\label{fig1} \includegraphics[width=0.45\textwidth]{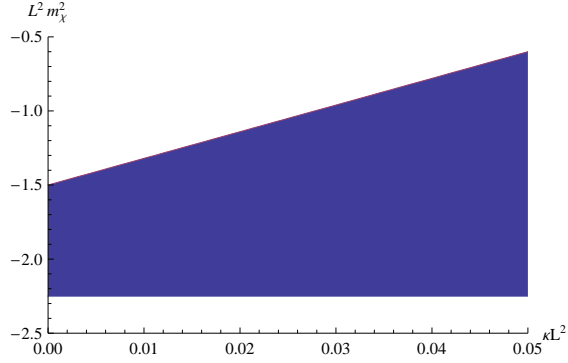}
\caption{BF bounds set by $AdS_4$ at UV and $AdS_2$ at IR, the scalar $\chi$ will condensate if $m_\chi^2L^2$ falls inside the shadow region.}
\end{figure}

We remark that the couple between scalar and Maxwell field contributes to the last term in the definition of effective mass, and apparently it depends on chemical potential.  The couple between two scalars does not enter due to a trivial $\psi_0$ in the $AdS_2$ background.  Since we are interested in the positive coupling constant $\kappa$, which can be tuned to shift the effective mass to be more negative.  In practice, for a neutral scalar to condensate, we ask the effective mass to satisfy $AdS_4$ Breitenlohner-Freedman (BF) bound\cite{Breitenlohner:1982bm} but violate that of $AdS_2$.  Therefore we are free to tune $\kappa$ such that
\begin{equation}
-\frac{9}{4} < m_\chi^2 L^2, \qquad (m_\chi^2-6\kappa \mu^2) L^2 \le -\frac{3}{2}. 
\end{equation}
We plot the admissible range of $\kappa$ and $m_\chi^2$ for condensate to happen in the Figure 1.  In particular, the equality holds for a critical $\kappa_c$ given some chosen mass\footnote{In the extremal limit, we are free to replace $\mu=\sqrt{3}$.},
\begin{equation}
\kappa_c = \frac{m_\chi^2}{18} + \frac{1}{12L^2},
\end{equation}
such that one may engineer a quantum phase transition where $T_c=0$ as that in \cite{Faulkner:2009wj}.  The equation (\ref{eqn_inner}) can be solved explicitly and the retarded Green function reads
\begin{eqnarray}
&&{\cal G}_k(\omega) = 2\nu_k e^{-i\pi \nu_k}\frac{\Gamma(-2\nu_k)\Gamma(\frac{1}{2}+\nu_k-iq_\chi \mu L_0)}{\Gamma(2\nu_k)\Gamma(\frac{1}{2}-\nu_k-iq_\chi \mu L_0)}(2\omega)^{2\nu_k},\nonumber\\
&&\nu_k \equiv \sqrt{m_{eff}^2 L_0^2-q_\chi^2\mu^2 L_0^2+\frac{1}{4}}.
\end{eqnarray}
Since the effect of nonminimal coupling only appears in the modification of effective mass, one expects the discussion in \cite{Faulkner:2009wj} still hold in our case.  To list a few:

\begin{enumerate}
	\item The low energy behavior of boundary system is uniquely determined by this IR analysis.  For example, the spectral function $Im_{} G_R(\omega,k) \propto \omega^{2\nu_k}$.
	\item For sufficient large $\kappa$, $\nu_k$ becomes pure imaginary and log-periodic behavior is expected and gapless excitation is responsible for this.
	\item Once a nonminimal coupled Fermion can be formulated in this background, one may have a model of non-Fermi liquids characterised by coupling $\kappa$. We will leave this for future study.
\end{enumerate}

%%%%%%%%%%%%%%%%%%%%%%%%%%%%%%%%%%%%%%%%%%%%%%%%
\section{Quantum criticality at Lifshitz point}
%%%%%%%%%%%%%%%%%%%%%%%%%%%%%%%%%%%%%%%%%%%%%%%%%

Now let us take a closer look at quantum critical point of generic Lifshitz scaling.  The equation of motion reads:
\begin{eqnarray}\label{eom_z}
&&-\partial_u^2\chi + \frac{(z+1)}{u}\partial_u\chi + \left[ k^2-(\frac{u}{L_0})^{2z-2}(\omega+q_\chi\sqrt{2-\frac{2}{z}}(\frac{L_0}{u})^z)^2+\frac{m_{eff}^2L_0^2}{u^2} \right]\chi=0,\nonumber\\
&&m_{eff}^2 \equiv m_\chi^2 - \frac{12(z-1)}{L^2(z+1)(z+2)}\left[ \kappa z - \frac{\eta}{m_\psi^2} \right].
\end{eqnarray}
At small $u$, the scalar $\chi$ behaves like
\begin{equation}
\chi \sim A(\omega) u^{1+\frac{z}{2}-\nu} + B(\omega) u^{1+\frac{z}{2}+\nu},\qquad \nu \equiv \sqrt{m_{eff}^2L_0^2-q_\chi^2(2-\frac{2}{z})L_0^2+(\frac{z+2}{2})^2}.
\end{equation}
By scaling invariance, one can argue the retarded Green function should scale like 
\begin{equation}
{\cal G}(\omega) \equiv \frac{B(\omega)}{A(\omega)} \propto \omega^{\frac{2\nu}{z}},\qquad 
\end{equation}
It is unclear whether the remaining part of $\cal G$ is obtainable for generic $z$ because analytic solution to equation (\ref{eom_z}) may not exist.  There is an exception for $z=2$.  One can find that, up to a normalized factor and phase:
\begin{eqnarray}\label{Retarded_Green}
&&{\cal G} \sim \frac{\Gamma(-\nu)\Gamma(\frac{1+\nu}{2}+i\delta)}{\Gamma(\nu)\Gamma(\frac{1-\nu}{2}-i\delta)} \omega^{\nu},\nonumber\\
&&\nu \equiv \sqrt{(m_{eff}^2-q_\chi^2)L_0^2+4}, \qquad \delta \equiv \frac{k^2}{4\omega}-\frac{q_\chi}{2}
\end{eqnarray}
This solvability mainly thanks to the integrability of equation (\ref{eom_z}) in the case of $z=2$, where it can be recasted into

\begin{equation}\label{ode_z2}
-\partial_u^2 \xi + \left[\frac{\nu^2}{u^2}-\frac{\omega^2}{L_0^2}u^2\right]\xi = -4\omega \delta\xi,
\end{equation}

with a field redefinition $\chi \equiv u^{3/2}\xi$.  This is nothing but a one-dimensional quantum mechanics of the Calogero particle with an inverse square potential in the harmonic trap.  A detail treatment for solving this is given in the appendix.

\begin{figure}[tbp]
\label{fig2} 
\includegraphics[width=0.45\textwidth]{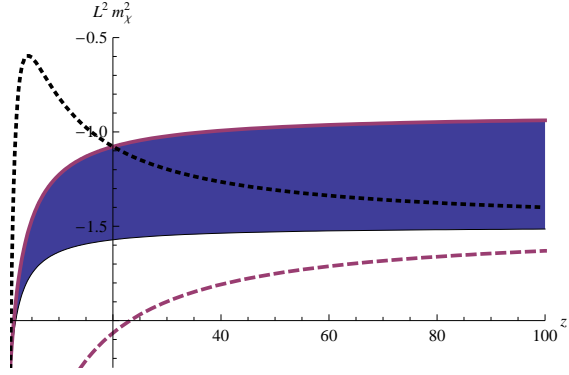}
\caption{A typical $\kappa=0.05$ will widen an additional window for condensate (shaded region).  We also reproduce the curves due to effect of negative $\eta$ (dotted black) and positive $\eta$ (dashed purple), below which the condensate develops.}
\end{figure}

\begin{figure}[tbp]
\label{fig3} 
\includegraphics[width=0.45\textwidth]{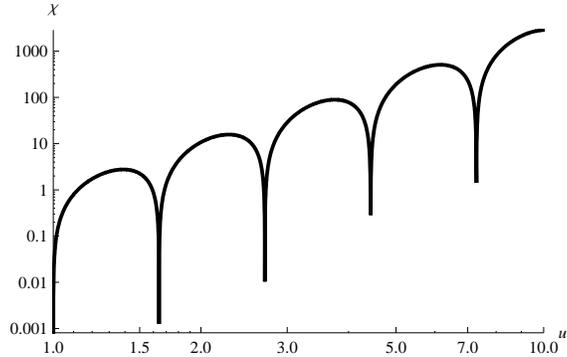}
\caption{A typical wavefunction for $|\chi|$, where the first four Efimov states are shown to correspond to those zeros at $u=1.643,2.791,4.438,7.293$ (The curve does not appear to hit the zero at those points due to limited number of sample points in the plot).  The second and higher Efimov states are higher excitation modes with one and more zero nodes.}
\end{figure}

\section{BKT phase transition at critical line}
To have condensate at IR, we ask the effective mass to violate the BF bound, that is
\begin{equation}
m_\chi^2L^2 \le \frac{12(z-1)}{(z+1)(z+2)}(\kappa z - \frac{\eta}{m_\psi^2})-\frac{3(z+2)}{2(z+1)}.
\end{equation}
In the Figure 2, we show that a shadow region is sandwiched by two curves, where the lower(upper) one associates to a BF bound with (non)zero $\kappa$.  This shows that nonminimal coupling enhances the instability and raises the BF bound for $m_\chi^2$.  In comparison, we also reproduce those BF bounds with positive and negative coupling $\eta$ but without $\kappa$\cite{Basu:2010fa}.  The quantum Lifshitz point corresponds to where the BF bound is about to be violated.  

Therefore, given masses of scalars and exponent $z$ there exists a critical line in the parameters space formed by $(\kappa, \eta)$.  For $\kappa z - \frac{\eta}{m_{\psi}^2}> \delta_c$, the BF bound is violated and  conformality is lost.  Following similar argument in \cite{Kaplan:2009kr,Faulkner:2009wj}, an infinite tower of IR scales is generated and associated with the infinite number of Efimov states: 
\begin{equation}
u_{IR}^{(n)} = u_{UV} \log(\frac{n\pi}{\sqrt{2(z-1)(\kappa z-\frac{\eta}{m_\psi^2}-\delta_c)}}), \qquad n = 1,2,\cdots.
\end{equation}
If we manage to turn on the temperature in this background, one may associate $u_{UV}$ to the scale set by $\mu$ and $u^{(1)}_{IR}$ to the scale set by a finite temperature $T_c$ as
\begin{equation}
T_c \sim (\frac{1}{u^{(1)}_{IR}})^z  \sim \mu \exp{(-\frac{z\pi}{\sqrt{2(z-1)(\kappa z-\frac{\eta}{m_\psi^2}-\delta_c)}})},
\end{equation}
In the Figure 3, we show that the first few IR scales out of infinite many, while the boundary condition of vanishing wavefunction $\chi$ is imposed at a chosen UV scale.  

In the Figure 4, it is also shown that the distance between UV scale and the first IR scale decreases with increasing $z$, implying that condensate is easier to form at larger $z$.  We observe that critical temperature rises up with increasing $\kappa-\kappa_c$ and plot it against various $z$ in the plot to the right.

We comment on some new features as follows: the positive $\kappa$ acts like a negative coupling $\eta$ in the case of $m_\psi^2 >0$, both seem to weaken the stability by decreasing its effective mass.  However, this similarity breaks down for large enough $z$, where the critical temperature is almost determined by $\kappa$ alone as follows:
\begin{equation}
T_c \sim \mu \exp(-\frac{\pi}{\sqrt{2\kappa}}),
\end{equation}
where the contribution from $\eta$ term is ignorable.  We remark that the minimal coupling limit $\kappa \to 0$ drives the critical $T_c \to 0$, which can be identified as the quantum critical point observed in pure $AdS_2$ background.

\begin{figure}[tbp]
\label{fig4} 
\includegraphics[width=0.45\textwidth]{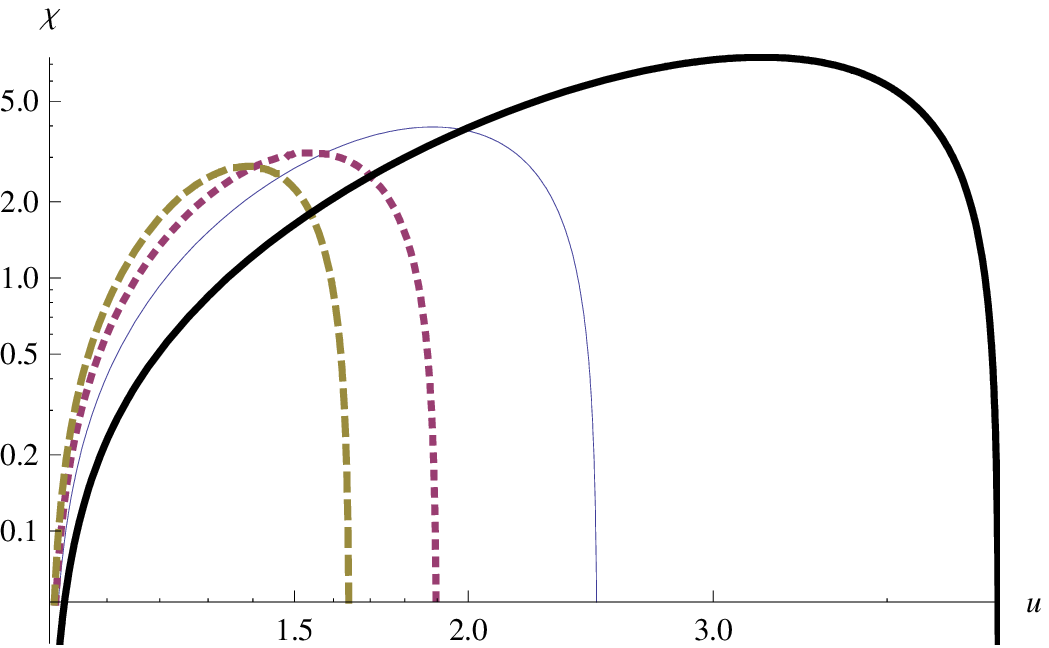} 
\includegraphics[width=0.45\textwidth]{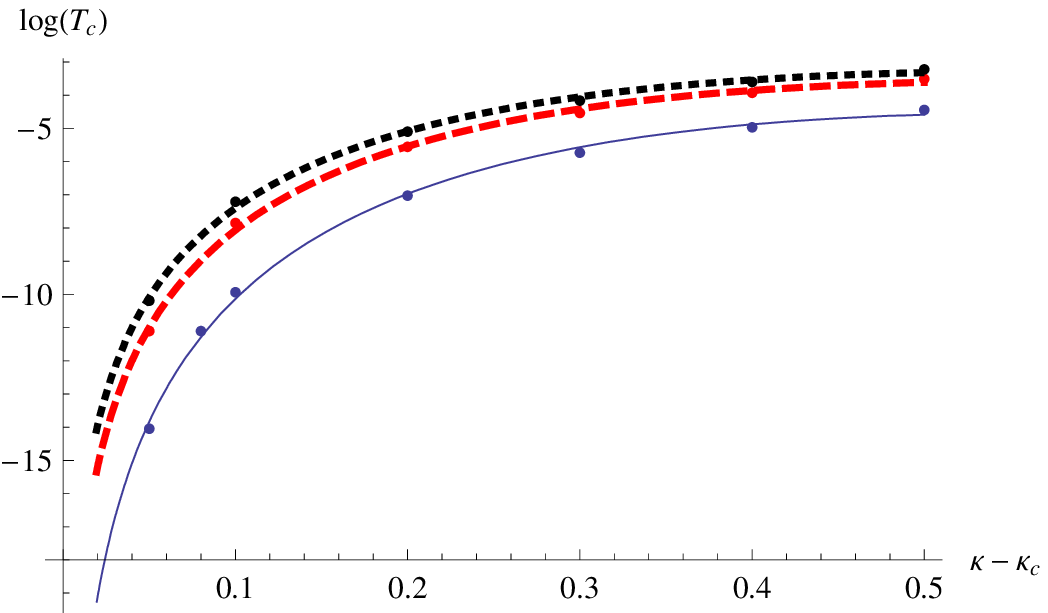}
\caption{To the left: IR scales against various $z$.  The zero of each $\chi$ at $u=1$ is chosen for the UV cutoff (by imposing the boundary condition) and another zero at larger $u$ can be regarded as the IR cutoff.  We have removed irrelevant IR scales set by higher Efimov states from the plot.  From right to left, the curves correspond to that of $z=2$(black thick), $z=3$(blue thin), $z=4$(red dotted) and $z=5$(green dashed).  The dynamically generated IR scales move toward the UV as $z$ increases, signaling a raise of critical temperatures.  Both axes are in the log scale.  To the right: The critical temperature against $\kappa-\kappa_c$ for $z=2$(blue), $z=5$(dashed red) and $z=20$(dotted black).  It is expected to reach quantum Lifshitz point at $\kappa = \kappa_c$.}
\end{figure}

\begin{figure}[tbp]
\label{fig5} 
\includegraphics[width=0.45\textwidth]{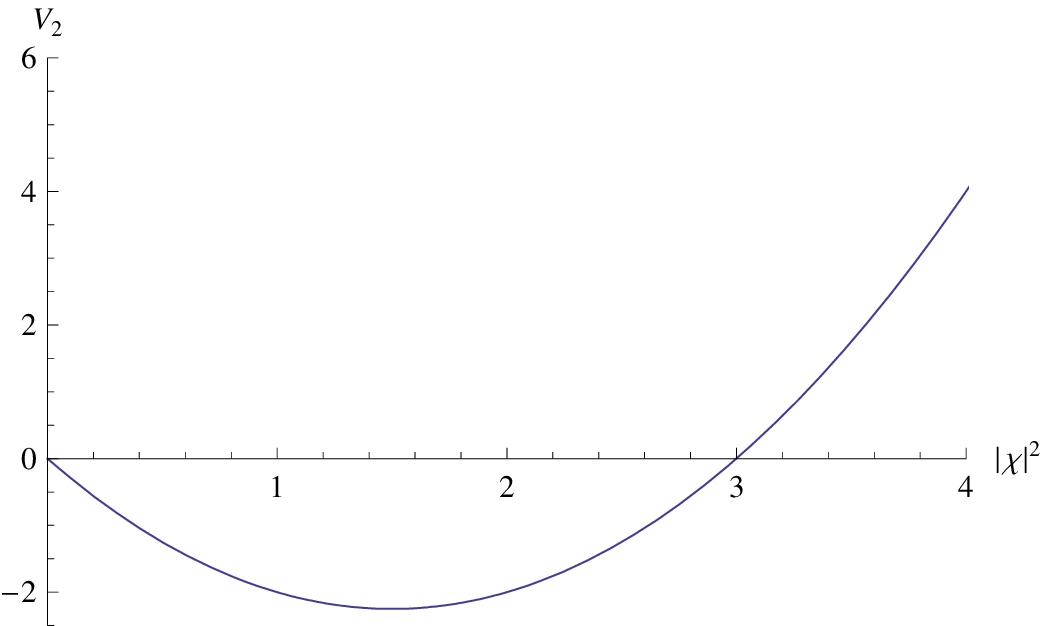} 
\includegraphics[width=0.45\textwidth]{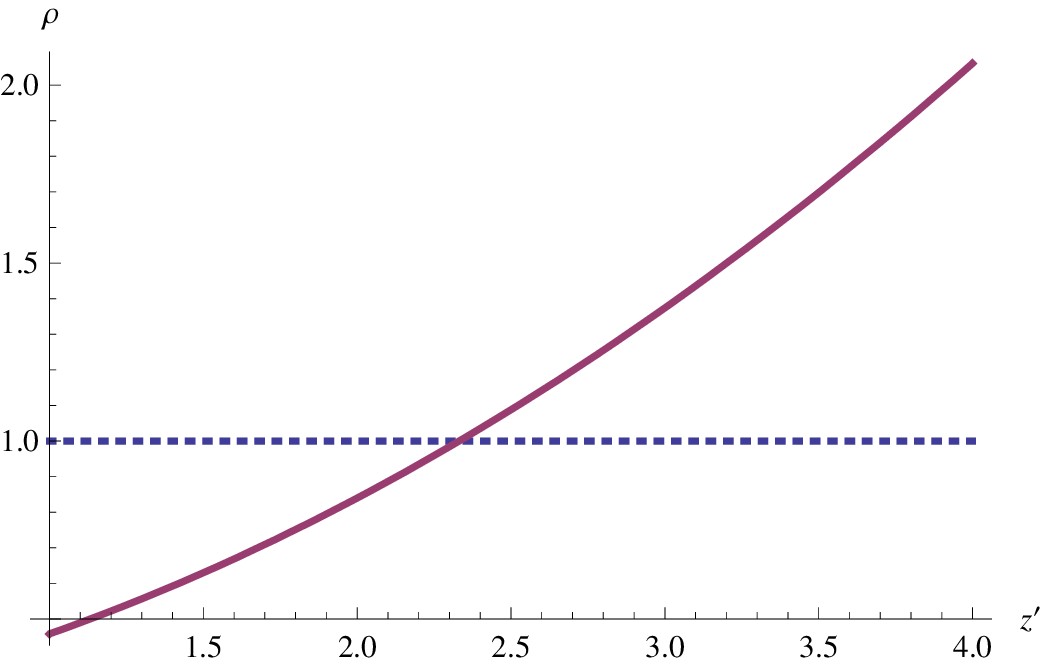}
\caption{To the left: A typical potential for $V_2(\chi)$.  At the minimum of potential, $|\chi|^2$ takes value $3/2$ at the IR cutoff.  To the right: A Plot of the ratio $\rho = (\tilde{L_0}/L_0)^2$ against $z'$ (thick red).  Notice for large enough $z'$, we have $\rho > 1$ (dashed blue), meaning that $\tilde{L_0}$ is larger than $L_0$.  For these plots, we have fixed the variables $z=2,q_\chi^2L_0^2=2,m_\chi^2L_0^2=-1.9,\kappa L_0^2=0.05, \lambda L_0^2 = 1$}
\end{figure}

\section{Beyond the probe}
To go beyond the probe limit, one starts to consider the back reaction from the $\chi$ field.  We will make the following assumptions, following similar arguments in \cite{Iqbal:2010eh}:

\begin{enumerate}
	\item We assume the existence of an IR cutoff point $u=u_0$, where $\chi$ field smoothly goes to a constant $\chi_0$.  This could be achieved by embedding a charged black hole and the horizon naturally introduces the cutoff\cite{Goldstein:2009cv}.  For our purpose here, we simply demand $\chi(u_0)= \chi_0$ and $\chi'(u_0)=0$ at the cutoff.
	  
	\item Nonlinear potential terms of higher power are necessary in addition to those in the equation (\ref{model}), in order to arrive at some physical ground state after receiving back reaction.   A simplest addition is to include a $|\chi|^4$ term.

	\item	After back reaction, we assume our background geometry still respects the Lifshitz scaling of some  critical exponent $z'$, which, however, is not necessary to be the same as the original $z$.

\end{enumerate}

Now we are ready to discuss the consequence derived from those assumptions.  Let us first investigate the trace of Einstein equation, including those parts with $\chi$ involved:
\begin{eqnarray}\label{trace}
&&R+ \frac{2(z^2+2z+3)}{L_0^2} = |D\chi|^2 + 2 V_1(\chi),\nonumber\\
&&V_1(\chi) \equiv  \left[ m_\chi^2 + \eta \frac{12}{m_\psi^2 L^2}\frac{z-1}{(z+1)(z+2)}\right] |\chi|^2 + \lambda |\chi|^4.
\end{eqnarray}
Notice that we have included the quartic terms with any $\lambda >0$.  
As what has been observed in \cite{Iqbal:2010eh}, it will reduce the effective $L_0$ providing that the right hand side of (\ref{trace}) is negative at IR cutoff.  However, if the variation of $z$ is admissible, it could increase $L_0$ instead.  To see this, we derive the effective radius of curvature for generic $z'$, denoting $\tilde{L}_0$, after receiving back reaction:
\begin{equation}
\frac{z'^2+2z'+3}{\tilde{L_0}^2} = \frac{z^2+2z+3}{L_0^2}-\left[ -q_\chi^2\frac{z-1}{z} + m_\chi^2 + \frac{2\eta}{m_\psi^2 L_0^2}(z-1) \right]|\chi_0|^2 - \lambda|\chi_0|^4.
\end{equation}
It is not difficult to see that $\tilde{L_0}$ in fact could increase if $z'$ is larger enough than $z$.  Interestingly, this in turns will either increase or decrease the background condensate $\psi$ thanks to its inversely proportional to $\tilde{L_0}$ as shown in the equation (\ref{sol_z}).  We plot a typical $V_2$ and the ratio $(\tilde{L_0}/L_0)^2$ in the Figure 5.

One should also investigate the equation of motion (\ref{eom_z}) around the IR cutoff:
\begin{eqnarray}
&&-u^2\partial_u^2\chi + L_0^2V_2'(\chi) = 0, \nonumber\\
&&V_2(\chi) \equiv \left[ -q_\chi^2 \frac{z-1}{z}+ m_\chi^2 - (z-1)\left( \kappa z - \frac{\eta}{m_\psi^2} \right) \right] |\chi|^2+\lambda |\chi|^4.
\end{eqnarray} 
To ensure the $\chi$ field sits right on the bottom of a concave-up potential at this point, we should demand $V_2'(\chi_0) = 0$ as well as 
$V_2''(\chi_0) > 0$.  This pins down to the following constraint:
\begin{equation}
-q_\chi^2 \frac{z-1}{z}+ m_\chi^2 - (z-1)( \kappa z - \frac{\eta}{m_\psi^2}) < 0
\end{equation}

We remark that potential $V_2$ includes the contribution from coupling $\kappa$ but $V_1$ does not thanks to the traceless condition of Maxwell field in four dimensions.

\begin{figure}[tbp]
\label{fig6} 
\includegraphics[width=0.45\textwidth]{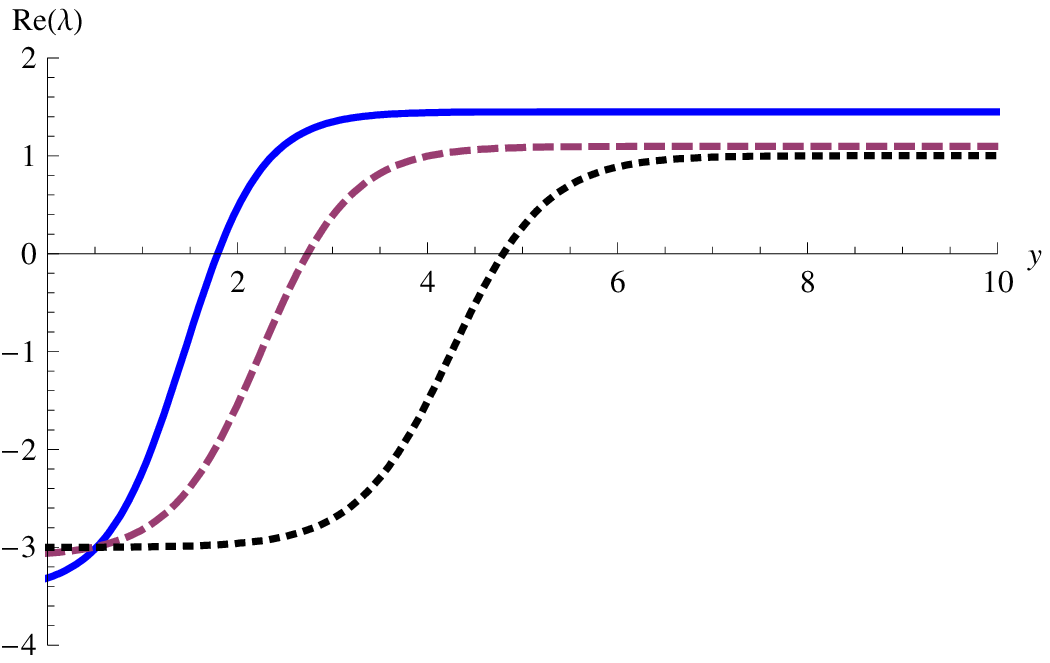} 
\includegraphics[width=0.45\textwidth]{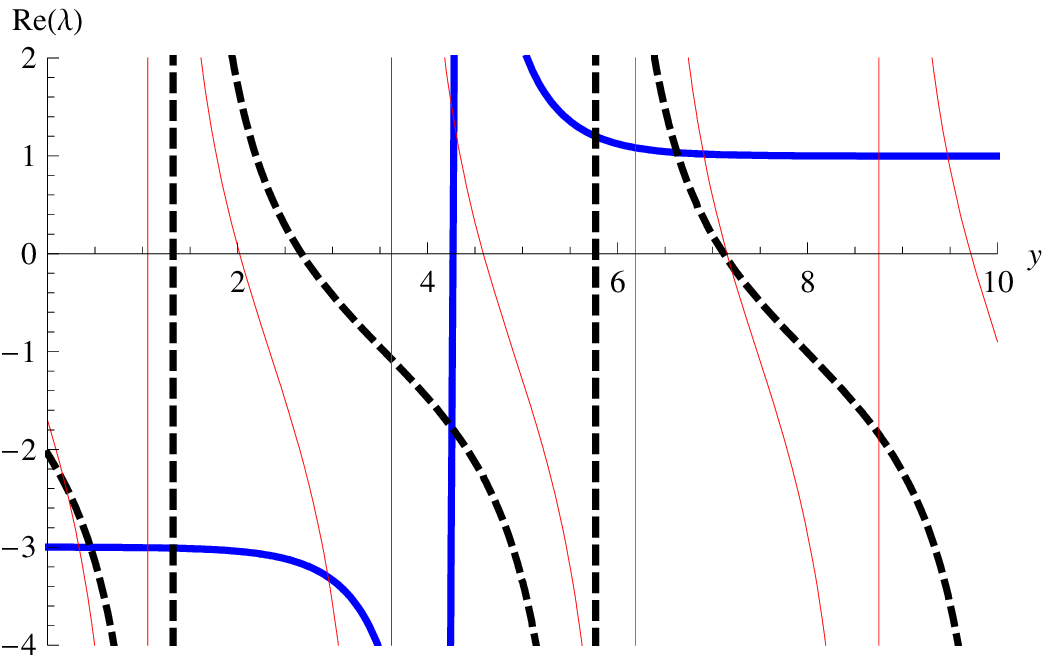}
\caption{To the left: A typical RG running for $\lambda$ for $\nu^2>0$, where nonminimal coupling $\kappa=0$(thick blue), $\kappa=0.2$(dashed red), $\kappa=0.249$(dotted black).  To the right: RG running shows periodic flows for $\kappa=0.251$(thick blue), $\kappa=1$(dashed black), $\kappa=1.5$(thin red).  In both plots, we have fine tuned the parameters to have critical $\kappa_c = 0.25$.}
\end{figure}

\section{Discussion}
Our model may be useful to describe a condensed matter system with two or multiple condensates, such as a two-band model in the superconductor.  Since we have taken one of two scalars to be small, our scenario is suitable to the window where a second condensate just begins to develop, while the first has been strong out there.  The above discussion regarding back reaction implies that the appearance of a second condensate may either enhance or suppress the first one through their direct coupling.  On the other hand, this back reaction may be seen as some sort of perturbation or deformation from the critical point, as recently discussed in \cite{Faulkner:2010gj}.

We have given some detail treatment for a Lifshiz system with critical exponent $z=2$ in the appendix, thanks to its integrability.  We showed that it can be translated into a one-dimensional quantum mechanics of Calogero particle in a harmonic trap potential and demonstrated that the RG running of its contact coupling shows periodic flowing once the unitarity is broken by overcritical $\kappa$.  For generic $z>2$, the differential equation (\ref{eom_z}) will include trap potential terms of higher order $O(u^z)$.  Since the contact potential, originally introduced for regularization, always dominates over the trap potential of any order in the region $u<u_0$, we expect that the same discussion for periodic RG running holds true for a Lifshitz system of higher $z$. In fact, this statement is confirmed in the section V by the appearance of an infinite tower of Efimov states in the bulk. 

%%%%%%%%%%%%%%%%%%%%%%%%%%%%%%%%%%%%%%%%%
\section*{Appendix}

In this appendix, we would like to take a closer look at a special case for $z=2$.  In particular, we will highlight the relation between inverse square potential and holographic RG flow, following the same treatment as found in \cite{Moroz:2009nm}.  One starts with the following differential equation, obtained from (\ref{ode_z2}) after a change of variable:  
\begin{eqnarray}
&&-\partial_u^2 \xi - V(u)\xi = - (k^2 - 2\omega q_\chi )\xi,\nonumber\\
&&V(u) = -\frac{\nu^2}{u^2}+\frac{\omega^2}{L_0^2}u^2.
\end{eqnarray}
In order to explore both regions of positive and negative $\nu^2$, we will analytically continue to a complex $u$ plane.  The above potential is ill-defined at $u=0$ for its inverse square potential.  One way to regularize it is to cut off the potential for $|u|< u_0$, and replace it by a $\delta-$function potential at $u=0$.  That is,
\begin{equation}
V(u) = -\frac{\lambda}{u_0}\delta(u),\qquad |u|< u_0,
\end{equation}
where $\lambda$ is a dimensionless contact coupling and its RG flow against the cutoff $u_0$ will be studied in the following.
The wavefunction can be exactly solved; it is given by the Parabolic Cylinder function for $|u|< u_0$ and the hypergeometric function outside the cutoff.  It is sufficient for us to work with the limit $u>0, y\equiv \sqrt{\omega}u \ll 1$, where $\xi(u)$ can be expanded as
\begin{equation}
\xi(y) = 
\left\{\begin{matrix}
(a + b y + O(y^2)) + D (a'+ b' y + O(y^2)), \quad 0< y <|y_0|\equiv \sqrt{\omega}u_0, \\ 
N(c_- y^{1/2-\nu} + c_+ y^{1/2+\nu}), \quad |y| > y_0.
\end{matrix}\right.
\end{equation}
where $a,a',b,b'$ are first two coefficients of Taylor expansion of the desired Parabolic Cylinder function, and $N$ is some normalization factor of wavefunction.  Their precise forms are irrelevant to our discussion here.  By observing the continuity of $\xi(y)$ and discontinuity of its first derivate at $y=0$, one can determine
\begin{equation}\label{D_lambda}
\lambda = 2 y_0\frac{-b-b'D}{a+a'D}.
\end{equation} 
Imposing the boundary condition of continuity of $\xi(y)$ and $\xi'(y)$ at $y=y_0$ for both expansions, one obtains the relations:
\begin{equation}\label{BC}
\left\{\begin{matrix}
a+Da' \approx N(c_-y_0^{1/2-\nu}+ c_+ y_0^{1/2+\nu}),\\
b+Db' \approx N\left[ c_-(1/2-\nu)y_0^{-1/2-\nu}+c_+(1/2+\nu)y_0^{-1/2+\nu}\right].
\end{matrix}\right.
\end{equation} 
Combining (\ref{D_lambda}) and (\ref{BC}), and introducing a running variable $t=-\ln y_0$, we arrive at the RG flow for $\lambda$
\begin{equation}
\lambda (t) = -1 + 2\nu \frac{e^{\nu t}-Ce^{-\nu t}}{e^{\nu t}+ Ce^{-\nu t}},\qquad C\equiv \frac{c_+}{c_-},
\end{equation}
where $C$ is nothing but retarded Green function ${\cal G}$ given in (\ref{Retarded_Green}).  The coupling $\lambda$ satisfies the general Riccati differential equation:
\begin{equation}
\partial_t \lambda = -\frac{1}{2}(\lambda + 1 - 2\nu)(\lambda + 1+ 2\nu).
\end{equation}
We plot it against various $\kappa$ in the Figure 6.  We remark that the periodic flows start to appear once the unitary bound is violated by $\kappa > \kappa_c$. 

%%%%%%%%%%%%%%%%%%%%%%%%%%%%%%%%%%%%%%%%%

\begin{acknowledgments}
The author would like to thank the hospitality of high energy theoretical group in the Caltech.  The author is grateful to useful discussion with Yutin Huang, Hirosi Ooguri, and Shang-Yu Wu. This work is supported in part by the Taiwan's National Science Council and National Center for Theoretical Sciences.
\end{acknowledgments}

%%%%%%%%%%%%%%%%%%%%%%%%%%%%%%%%%%%%%%%%%%%%%%%%%%%%%%%%%%%%%%%

\bibliography{apssamp}% Produces the bibliography via BibTeX.

\end{document}